\documentclass[12pt,thmsa]{article}
\usepackage{amsfonts}
\usepackage{latexsym}
\usepackage{amsmath}

\begin{document}

\author{Carlos N. Kozameh$^{1}$ and Ezra T. Newman$^{2}$ \\
$^{1}$FaMAF, Universidad Nacional de C\'{o}rdoba, 5000 C\'{o}rdoba, Argentina%
\\
$^{2}$Dept of Physics and Astronomy, Univ. of Pittsburgh, Pgh, PA 15260}
\title{The Large Footprints of H-Space on Asymptotically Flat Space-Times}
\date{March 27, 2005 }
\maketitle

\begin{abstract}
We show that certain structures defined on the complex four dimensional space known as H-Space have considerable
relevance for its closely associated asymptotically flat real physical space-time. More specifically for every
complex analytic curve on the H-space there is an asymptotically shear-free null geodesic congruence in the
physical space-time. There are specific geometric structures that allow this world-line to be chosen in a unique
canonical fashion giving it physical meaning and significance.
\end{abstract}

\section{Introduction}

About twenty-five years ago the theory of H-space\cite{H-space1,H-space2}, was developed from considerations of
asymptotically flat solutions to the Einstein equations. Given an arbitrary asymptotically flat space-time the
question raised was: could one find, in general, in the neighborhood of null infinity, null surfaces that were
shear free. The answer, in the general situation, was no, one could not. However, if the physical space-time was
real analytic and was allowed to be complexified, one could show that there existed a four-complex parameter
family of complex null surfaces that were asymptotically shear-free. These four parameters defined the
coordinates of H-space.

It turned out that H-space naturally possessed a complex (holomorphic)
metric that was both Ricci flat and half-conformally flat. Though H-space
had many fascinating properties, the fact that it was intrinsically complex
severely limited its physical relevance.

Very recently, however, we discovered that if our original question of
finding complex shear-free-null surfaces was changed to the question of; can
we find \textit{real null geodesic congruences} that were again \textit{\
asymptotically} shear-free but now \textit{were not necessarily surface
forming}, i.e., that they could possess \textit{twist}, the answer was yes.

It turns out that every such congruence is determined by the choice of a
complex (holomorphic) world-line in H-space and conversely, every complex
H-space world-line determines an asymptotically shear-free null geodesic
congruence. In other words, given the parametric form of an H-space curve
\[
z^{a}=\xi ^{a}(\tau )
\]
with $z^{a}$ points of H-space and $\tau $ a complex parameter, there is
determined an asymptotically shear-free congruence that, in general, is
twisting.

[In the special case when the space-time possesses a Bondi shear $\sigma ,$
that is purely of `electric' type the H-space is flat and can be identified
with complex Minkowski space. In this case, real world-lines determine
\textit{real} asymptotically shear-free \textit{null surfaces}.]

In Section II, we show how the H-space world-lines determine the
asymptotically shear-free congruences, while in Section III we list some
special cases and some applications of these observations - but described in
detail elsewhere.

\section{Derivations and Proofs}

Starting with an asymptotically flat space-time, we begin with Bondi
coordinates, ($u,\zeta ,\overline{\zeta }),$ on future null infinity, ($%
\frak{I}^{+}$),  and the associated Bondi tetrad system ($%
l_{B}^{a},m_{B}^{a},\overline{m}_{B}^{a},n_{B}^{a}$) with $n_{B}^{a}$
tangent to a generator of $\frak{I}^{+}$ and $l_{B}^{a}$ tangent to a
generator of the Bondi null surface, $u=const.$ The Bondi asymptotic shear
is an arbitrary, but given, spin-weight 2 function, $\sigma (u,\zeta ,%
\overline{\zeta }).$ If we perform an asymptotic null rotation around $%
n_{B}^{a}$ given by

\begin{eqnarray}
l^{*a} &=&l_{B}^{a}+b\overline{m}_{B}^{a}+\overline{b}m_{B}^{a}+b\overline{b}%
n_{B}^{a},  \label{NullRot} \\
m^{*a} &=&m_{B}^{a}+bn_{B}^{a},  \nonumber \\
n^{*a} &=&n_{B}^{a},  \nonumber \\
b &=&-L/r+O(r^{-2})  \nonumber
\end{eqnarray}
we can ask what is the asymptotic shear $\sigma ^{*}(u,\zeta ,\overline{%
\zeta })$ of the new null congruence with tangent vector $l^{*a}.$ After a
lengthy calculation\cite{Aronson}\textbf{,} the new shear of the new null
vector $l^{*a}$ is related to the old one by
\[
\sigma ^{*}(u,\zeta ,\overline{\zeta })=\sigma (u,\zeta ,\overline{\zeta })-%
\partial L-L\; L,_{u}.
\]

Thus, to make the new shear vanish, the function $L(u,\zeta ,\overline{\zeta
})$ must satisfy the differential equation
\begin{equation}
\partial L+L \; L,_{u}=\sigma (u,\zeta ,\overline{\zeta }).  \label{ShearFree}
\end{equation}
It is this equation, which plays the major role here, that we must analyze.

First we define a new variable, $\tau ,$ and its inversion
\begin{eqnarray}
\tau &=&T(u,\zeta ,\overline{\zeta })  \label{tau} \\
u &=&X(\tau ,\zeta ,\overline{\zeta })  \label{X}
\end{eqnarray}
by its relationship with $L$ via
\begin{equation}
L=-\frac{\partial T}{T,_{u}}.  \label{L&tau}
\end{equation}
If this form is immediately substituted into Eq.(\ref{ShearFree}), there is
no immediate simplification. If however, $\tau $ is given implicitly by Eq.(%
\ref{X}) rather than by Eq.(\ref{tau}), a \textit{huge simplification occurs}%
.

From Eq.(\ref{X}), by implicit differentiation, we have
\begin{eqnarray}
0 &=&\partial_{(\tau )}X(\tau ,\zeta ,\overline{\zeta })+X^{\,\prime }%
\partial T  \label{implicit1} \\
1 &=&X^{\,\prime }(\tau ,\zeta ,\overline{\zeta })T,_{u}  \label{implicit2}
\end{eqnarray}
where the prime, ($^{\prime }$), implies the $\tau $ derivative and $\partial _{(\tau )}$ means $\partial$
holding $\tau $ constant. Using Eqs.(\ref{implicit1}) and (\ref{implicit2}) in Eq.(\ref{L&tau}) we have
\begin{eqnarray}
L(u,\zeta ,\overline{\zeta }) &=&\partial_{(\tau )}X(\tau ,\zeta ,%
\overline{\zeta })  \label{L&tau2} \\
u &=&X(\tau ,\zeta ,\overline{\zeta }).  \nonumber
\end{eqnarray}
Finally, again using Eqs.(\ref{implicit1}) and (\ref{implicit2}), with Eq.(%
\ref{L&tau2}), we find that Eq.(\ref{ShearFree}) becomes
\begin{equation}
\partial_{(\tau )}^{2}X(\tau ,\zeta ,\overline{\zeta })=\sigma (X,\zeta ,%
\overline{\zeta }).  \label{Hspace}
\end{equation}
Now observing that this could be written without any reference to $\tau $ as
\begin{equation}
\partial^{2}X(\zeta ,\overline{\zeta })=\sigma (X,\zeta ,\overline{\zeta }%
),  \label{Hspace2}
\end{equation}
it is seen to be the well-known `good-cut' equation\cite{H-space1,H-space2}.
Its solutions are know\cite{H-space1} to depend on four complex parameters, $%
z^{a},$ i.e., it can be written as $X=$ $X(z^{a},\zeta ,\overline{\zeta }).$
To restore the $\tau $ dependence all we must do is replace $z^{a}$ by
\begin{equation}
z^{a}=\xi ^{a}(\tau ),  \label{world-line}
\end{equation}
i.e., each solution is determined by a complex analytic world-line in
H-space as was claimed in the introduction. The general solution for $%
L(u,\zeta ,\overline{\zeta })$ is thus given, implicitely, by
\begin{eqnarray}
L(u,\zeta ,\overline{\zeta }) &=&\partial_{(\tau )}X(\xi ^{a}(\tau
),\zeta ,\overline{\zeta })  \label{L2} \\
u &=&X(\xi ^{a}(\tau ),\zeta ,\overline{\zeta }).  \label{X2}
\end{eqnarray}

Since we appear to be mixing real and complex valued functions, it is
worthwhile to briefly discuss this issue. We have assumed from the beginning
that we were dealing with a real analytic space-time (i.e., with a real
analytic metric tensor) so that the real coordinates could, \textit{if
needed, }be extended into the complex. The $u$ and $r$ could take complex
values and the complex conjugate stereographic pair ($\zeta ,\overline{\zeta
})$ could be freed from each other. Considering both $u$ and $\tau $ to be
complex, makes Eq.(\ref{X2}) a meaningful complex relationship that is not,
in general, a real analytic relationship. Nevertheless we are interested
only in real values for $u$. Assuming, of course, the invertibility of (\ref
{X2}), we take the values of $u$ as real and find the associated values of $%
\tau $ as ($\zeta ,\overline{\zeta })$ range over the real sphere, thus
mapping the sphere into the complex $\tau $-plane. For a real Bondi cut of $%
\frak{I}^{+}$, $u=const$, there are a spheres worth of points in H-space so
that for each of these points there is a null ray, given by Eq.(\ref{L2}),
that intersects the cut. Hence from a real point of view the complex
world-line becomes a world-tube, $S^{2}$x$R,$ in H-space. Every point on the
tube corresponds to a real null direction on $\frak{I}^{+}$. [There is the
degenerate situation when the tube collapses into a real world-line that
occurs when the H-space is flat and a real analytic world-line is chosen.]

Summarizing, we have the following: Beginning with an asymptotically flat
space-time with given Bondi shear $\sigma (u,\zeta ,\overline{\zeta }),$ we
construct (in principle) the associated H-space and chose on it a complex
analytic world-line. This in turn produces the solution Eqs.(\ref{L2}) and (%
\ref{X2}) and the new null direction field $l^{*a}.$

\textbf{Remark 1}
We point out that at each point, ($u,\zeta ,\overline{\zeta }),$ of $\frak{I}%
^{+}$ there is a past null cone with its celestial sphere. The variable $%
L(u,\zeta ,\overline{\zeta })$ is a natural complex stereographic coordinate
on that sphere so that the solution defines a null shear-free direction
field on $\frak{I}^{+}$ described either by the angle field $L(u,\zeta ,%
\overline{\zeta })$ or equivalently by the null vector field $l^{*a}.$
Though we have used a given Bondi slicing for the description of this angle
field, it is a geometric structure that is independent of the method of
description.

\section{Discussion}

We have seen in the previous section that all real asymptotically shear-free
null geodesic congruences can be described in terms of analytic curves in
the complex H-space. The question immediately arises; what use are these
congruences in either electromagnetic theory or in general relativity?
Though we will not discuss here these uses in any detail, there are a wide
variety of applications that will be described elsewhere. Here we will
simply outline them.

But before doing so, we discuss the important special case of asymptotically
flat space-times with a Bondi shear that is purely of `electric type'. The
term `electric type' means that the shear can be written as
\begin{equation}
\sigma (u,\zeta ,\overline{\zeta })=\partial^{2}S(u,\zeta ,\overline{%
\zeta })  \label{electrictype}
\end{equation}
with $S(u,\zeta ,\overline{\zeta })$ a \textit{real }spin-weight zero and
conformal weight one function. If an 'electric type' shear is used in Eq.(%
\ref{Hspace2}), i.e.,
\[
\partial^{2}X=\partial_{(X)}^{2}S(X,\zeta ,\overline{\zeta }),
\]
one sees that any real analytic solution $X=X(z^{a},\zeta ,\overline{\zeta }%
),$ also satisfies the equation for $\overline{\text{H}}$-spaces,
\[
\overline{\partial}^{2}X=\overline{\partial}_{(X)}^{2}S(X,\zeta ,%
\overline{\zeta }).
\]
This in turn means that we have simultaneously both an H-space and $%
\overline{\text{H}}$-space so that it is Ricci-flat and has both the
self-dual and anti-self-dual Weyl tensor vanishing, i.e., it is flat.
Coordinates can be adapted so that the metric is $g=\eta _{ab}dz^{a}dz^{b}$
and the world-lines of the previous section are complex world-lines in
Minkowski space, $z^{a}=\xi ^{a}(\tau ).$ If one chooses a world-line so
that $\xi ^{a}(\tau )$ is real analytic then the null geodesic congruence
obtained from $l^{*a}$ is surface forming.

Returning to the issue of what, if any, are the applications for these
asymptotic shear-free congruences, we mention several. The essential idea is
to find, from the existing geometric structure of the space-time (or from
the Maxwell fields), a means of uniquely choosing the world-line $z^{a}=\xi
^{a}(\tau )$ and then trying to give it a physical meaning.

1. If the physical space-time is flat with an asymptotically vanishing
Maxwell field, the Maxwell field itself contains certain structures that
allow a unique choice of $\xi ^{a}(\tau )$ that can be interpreted as a
\textit{complex center of charge} from which one obtains a geometric
definition of the electric and magnetic dipole moments. For example, if the
Maxwell field has a principle null vector field that is shear-free or even
asymptotically shear-free, then choosing it as the asymptotic field $l^{*a}$
determines $L$($u,\zeta ,\overline{\zeta })$ and the world-line $\xi
^{a}(\tau ).$ If the principle null vector field is twist free then the $\xi
^{a}(\tau )$ is a real world-line in real Minkowski space but if it has
twist then the world-line is complex. In the first case there will be no
magnetic dipole while in the second there will be a magnetic dipole. In the
special case of a Lienard-Wiechert field the physical source world-line is
our \textit{center of charge line\cite{ShearFreeMax}. }

2. If the space-time is algebraically special then there is already a
`degenerate' null geodesic congruence that is already totally shear free. We
can then find the unique $\xi ^{a}(\tau )$ and the unique $L$($u,\zeta ,%
\overline{\zeta })$ so that they coincide with those of the degenerate
congruence. One then finds `equations of motion' for the $\xi ^{a}(\tau ).$
A special case of this is the Robinson-Trautman type II metrics where the
world-line $\xi ^{a}(\tau )$ is a real line in Minkowski space and the
congruence is surface forming. In the case of the twisting algebraically
special metrics one gets evolution equations for the Pauli-Lubanski spin
vector identified from the imaginary part of $\xi ^{a}(\tau ).$

3. In the more general situation there are other geometric means, (based on
a very much weakened version of the idea of a principle null vector) for
uniquely choosing the $\xi ^{a}(\tau ).$ This world-line can be identified
or associated with a \textit{complex center of mass} that allows an
asymptotic definition of both orbital and spin angular momentum and an
asymptotic meaning to ``equations of motion''.

4. For the asymptotically flat Einstein-Maxwell fields, there will, in
general, be two different world-lines: a \textit{complex center of charge}
and a \textit{complex center of mass world-line. }If these two world-line
happen to coincide then the asymptotic system (appears, from work in
progress) to have a gyromagnetic ratio of the Dirac electron, i.e., $g=2.$

4. The construction of the Kerr and the charged Kerr metrics, based on a
complex world-line in Minkowski space, are special cases of the issues
discussed here\cite{gyro}.

Though there are still many questions concerning this subject that remain to
be clarified, nevertheless the observation that there is a large imprint of
H-space on real asymptotically flat space-time unifies many disparate ideas
and appears to open fresh research avenues.

\section{Acknowledgments}

This material is based upon work (partially) supported by the National
Science Foundation under Grant No. PHY-0244513. Any opinions, findings, and
conclusions or recommendations expressed in this material are those of the
authors and do not necessarily reflect the views of the National Science.
E.T.N. thanks the NSF for this support. C.N.K. would like to thank CONICET
for support.

\end{document}